\documentclass[ reprint,superscriptaddress,showpacs, amsmath,amssymb, aps, prl,floatfix,]{revtex4-1}

\usepackage{color}
\usepackage{graphicx}
\usepackage{dcolumn}
\usepackage{bm}
\usepackage{subfigure}
\usepackage{multirow}
\usepackage{braket}
\newcommand{\putat}[3]{\begin{picture}(0,0)(0,0)\put(#1,#2){#3}\end{picture}}

\newcommand{\BABARPubYear}       {16}
\newcommand{\BABARPubNumber}     {007}
\newcommand{\SLACPubNumber} {17118}

\input{babarsym}

\long\def\inst#1{\par\nobreak\kern 4pt\nobreak
    {\it #1}\par\vskip 10pt plus 3pt minus 3pt}
    
\def\finalsys {\ensuremath{12.9}\xspace}
\begin{document}

\preprint{\babar-PUB-\BABARPubYear/\BABARPubNumber} 
\preprint{SLAC-PUB-\SLACPubNumber} 

\onecolumngrid
\hbox to \hsize{
\vbox{
\begin{flushleft}
\babar-PUB-\BABARPubYear/\BABARPubNumber\\
SLAC-PUB-\SLACPubNumber\\
\end{flushleft}
\vspace{\baselineskip}
}
\hfill
}
\vspace{-\baselineskip}

\title{
\large \bfseries \boldmath
Measurement of the $D^{*}(2010)^{+} - D^+$ mass difference
}

%
\author{J.~P.~Lees}
\author{V.~Poireau}
\author{V.~Tisserand}
\affiliation{Laboratoire d'Annecy-le-Vieux de Physique des Particules (LAPP), Universit\'e de Savoie, CNRS/IN2P3,  F-74941 Annecy-Le-Vieux, France}
\author{E.~Grauges}
\affiliation{Universitat de Barcelona, Facultat de Fisica, Departament ECM, E-08028 Barcelona, Spain }
\author{A.~Palano}
\affiliation{INFN Sezione di Bari and Dipartimento di Fisica, Universit\`a di Bari, I-70126 Bari, Italy }
\author{G.~Eigen}
\affiliation{University of Bergen, Institute of Physics, N-5007 Bergen, Norway }
\author{D.~N.~Brown}
\author{Yu.~G.~Kolomensky}
\affiliation{Lawrence Berkeley National Laboratory and University of California, Berkeley, California 94720, USA }
\author{M.~Fritsch}
\author{H.~Koch}
\author{T.~Schroeder}
\affiliation{Ruhr Universit\"at Bochum, Institut f\"ur Experimentalphysik 1, D-44780 Bochum, Germany }
% ...
\collaboration{The \babar\ Collaboration}
\noaffiliation

\begin{abstract}
    We measure the mass difference, $\Delta m_+$, between the $D^{*}(2010)^+$ and the $D^+$ using the decay chain $D^{*}(2010)^+\to \Dp \piz$ with $\Dp\to \Km\pip\pip$. The data were recorded with the \babar detector at center-of-mass energies at and near the $\Upsilon(4S)$ resonance, and correspond to an integrated luminosity of approximately $468 \invfb$. We measure 
    $\Delta m_+ = \left(140\,601.0 \pm 6.8[{\rm stat}] \pm \finalsys[{\rm syst}]\right) \kev$. 
    We combine this result with a previous \babar measurement of $\Delta m_0\equiv m(D^{*}(2010)^+) - m (\Dz)$ to obtain 
    $\Delta m_D =  m(\Dp) - m(\Dz) = \left(4\,824.9 \pm 6.8[{\rm stat}] \pm \finalsys[{\rm syst}]\right)$~\kev. 
    These results are
compatible with and     
 approximately five times more precise than the Particle Data Group averages.
\end{abstract}
\pacs{13.20.Fc, 13.25Ft, 14.40.Lb, 12.38.Gc, 12.38.Qk, 12.39.Ki, 12.39.Pn}

\maketitle                                                                             

\setcounter{footnote}{0}
The difference
between the masses of the $ D^0 $ and $ D^+ $ mesons~\footnote{charge conjugation is
implied throughout this paper.}, 
$ \Delta m_D  \equiv  m(D^+) - m(D^0) $,  
is a key ingredient constraining calculations of symmetry
breaking due to differing $ u $ and
$ d $ quark masses
and electromagnetic interactions in 
the frameworks of chiral perturbation theory~\cite{Goity:2007fu}
 and lattice QCD~\cite{Horsley:2013qka}. 
 Its value is reported by the Particle Data Group (PDG)~\cite{pdg} 
 to be  $\Delta m_D = (4.77 \pm 0.08) $~\mev. 
The most precise direct measurement, 
reported by the LHCb Collaboration, is 
$ \Delta m_D = (4.76 \pm 0.12 \pm 0.07) $~\mev~\cite{Aaij:2013uaa}. 
This was found by comparing the invariant mass distributions of
$ D^0 \to K^- K^+ \pi^- \pi^+ $ and $ D^+ \to K^- K^+ \pi^+ $
decays. 
A more powerful constraint comes from the difference of measured
$ D^{*+} \to D^+ \pi^0 $ 
and $ D^{*+} \to D^0 \pi^+ $  mass difference distributions. 
CLEO has previously reported 
$\Delta m_+ \equiv m(D^{*}(2010)^+) - m(\Dp) = (140.64 \pm 0.08 \pm 0.06)$~\mev using
the decay chain $\Dstarp \to \Dp\piz$ with $\Dp\to\Km\pip\pip$~\cite{Bortoletto:1992rj}. 
In the present paper we report a new measurement
of $ \Delta m_+ $ and
combine it with our previously measured
$\Dstarp\to \Dz\pip$ mass difference~\cite{dstd0prl,prdversion}, 
$\Delta m_0 \equiv m(D^{*}(2010)^+) - m(\Dz)$, using two decay modes $\Dz\to \Km\pip$ and $\Dz \to \Km\pip\pim\pip$, to determine
$ \Delta m_D \equiv \Delta m_0 - \Delta m_+ $ 
with very high precision. 

This analysis is based on a data set corresponding to an integrated 
luminosity of approximately 468~\invfb recorded at, and 40 \mev below, 
the $\Upsilon(4S)$ resonance~\cite{Lees2013203}. 
The data were collected with the \babar detector at the \pep2\ asymmetric 
energy \epem\ collider, located at the SLAC National Accelerator Laboratory. 
The \babar detector is described in detail 
elsewhere~\cite{ref:babar,ref:nim_update}. 
The momenta of charged particles are measured with a combination of a 
cylindrical drift chamber (DCH) and a 5-layer silicon vertex tracker (SVT), 
both operating within the $1.5$ T magnetic field of a superconducting solenoid. 
Information from a ring-imaging Cherenkov detector is combined with specific 
ionization ($\dedx$) measurements from the SVT and DCH to identify charged 
kaon and pion candidates. 
Electrons are identified, and photons from $\piz$ decays are measured, with a CsI(Tl) 
electromagnetic calorimeter (EMC). 
The return yoke of the superconducting coil is instrumented with tracking 
chambers for the identification of muons. 

We study the $\Dstarp\to \Dp\piz$ transition, using the $\Dp \to \Km\pip\pip$
decay mode, to determine the difference between the $\Dstarp$ and $\Dp$ masses $\Delta m_+$. 
To extract $\Delta m_+$, we fit the distribution of the difference between
the reconstructed $\Dstarp$ and $\Dp$ masses, $\Delta m$. 
The signal component in the $\Delta m$ fit
 is a resolution function determined from 
our Monte Carlo (MC) simulation of the detector response, 
while the contaminations from background are accounted for 
by a threshold function.

We suppress combinatorial backgrounds, and backgrounds with $\Dstarp$ candidates from $B$ decays, 
by requiring $D^{*+}$ mesons produced in $e^+e^- \to c \bar{c}$ reactions 
to have momenta in the \epem center-of-mass frame greater than $ 3.0 \gev$. 
Decays 
$\Dstarp\to \Dz\pip$ with
$ D^0 \to K^- \pi^+ \pi^0 $ create backgrounds  
when the  $ \pi^+ $ daughter of the $\Dstarp\to \Dz\pip$  decay 
replaces the $\piz$ in the $\Dz$ decay
by mistake and the two have similar momenta.
To mitigate this problem, 
events 
are rejected if $m(\Km\pip\pip\piz) - m(\Km\pip\piz) < 160$~\mev 
for either of the two $\pip$. 
The value of 160~\mev is chosen to be very conservative in terms of
removing $\Dstarp\to \Dz\pip$ decays \cite{dstd0prl,prdversion} 
and causes almost no loss of
signal.
The decay chain is fitted subject to
geometric constraints at the $\Dstarp$ production vertex and  
the $\Dp$ decay vertex,  and to a kinematic constraint that the $\Dp$ 
laboratory momentum points back to the 
luminous region whose horizontal, vertical, and longitudinal 
RMS dimensions are about
6, 9, and 120 $ \mu $m, respectively 
\cite{ref:babar}.
The $\chi^2$ $p$-value from the fit is required to be
greater than 0.1\%. 

The ``slow pion'' from $\Dstarp$ decay, denoted as $\pi_s^0$,
has a typical laboratory momentum 
of 300~\mev. 
All photons from $\pi^0_s$ decays have energies below 500~\mev.
Their energy resolution is $ \sigma_E/E \sim7\%$ , and angular resolutions
 are  $  \sigma_\theta$ and $\sigma_\phi $  $ \sim 10 $ mr 
where  
the resolutions 
are measured with large uncertainties.
In the $\pi_s^0\to\gamma\gamma$ reconstruction, 
we first require both photon energies to be above 60~\mev, the total energy 
to be greater than 200~\mev, and the diphoton invariant mass to be between 
120~\mev and 150~\mev (approximately $\pm 2.5 \sigma$ around the nominal $\piz$ mass~\cite{pdg}). 
After the selection, each photon pair is kinematically fitted to the hypothesis of a $\piz$ originating from the event primary vertex, 
 and with the diphoton mass constrained to the nominal $\piz$ mass. This greatly improves the
 reconstructed $\piz$ momentum resolution, and 
therefore the $\Delta m$ resolution. The $\piz$ relative momentum resolution after the kinematic fit is $\sigma_p/p\sim$3\%; this is still considerably worse than the approximately 0.5\% $\Dp$ relative momentum resolution.  

Our MC simulation attempts to track run-by-run variations in 
detector response.
The standard MC energy calibration method that accounts for energy loss in the
EMC differs from that used with real data.
This results in a reconstructed $\piz$ mass ($m_{\gamma\gamma}$) peak 
in MC events that peaks about 0.5~\mev below the nominal mass for low energy
$ \pi^0 $s.
In contrast, the $m_{\gamma\gamma}$ peak value from the calibrated data events 
generally coincides with the nominal value. 
Therefore, we approximate the 
neutral energy correction algorithm used in data by 
rescaling
the reconstructed photon energies in MC 
events by factors 
depending on 
photon energy and data-taking period~\cite{ref:nim_update}. 
While this improves the  
data-MC agreement, 
the reconstructed $\piz$ momentum in MC events 
    remains slightly biased when compared with its generated
    value.
To account for this bias, we also rescale
 the \piz momentum in each MC 
 event by  
 approximately 0.2\%, 
depending on the diphoton opening angle.    
In addition to improving the data-MC agreement 
in peak positions and shapes of the
background-subtracted $m_{\gamma\gamma}$ distributions, these MC corrections
substantially improve the agreement in kinematic distributions, 
as described below.

\begin{figure}[ht]
\centering
\includegraphics[width=0.9\linewidth]{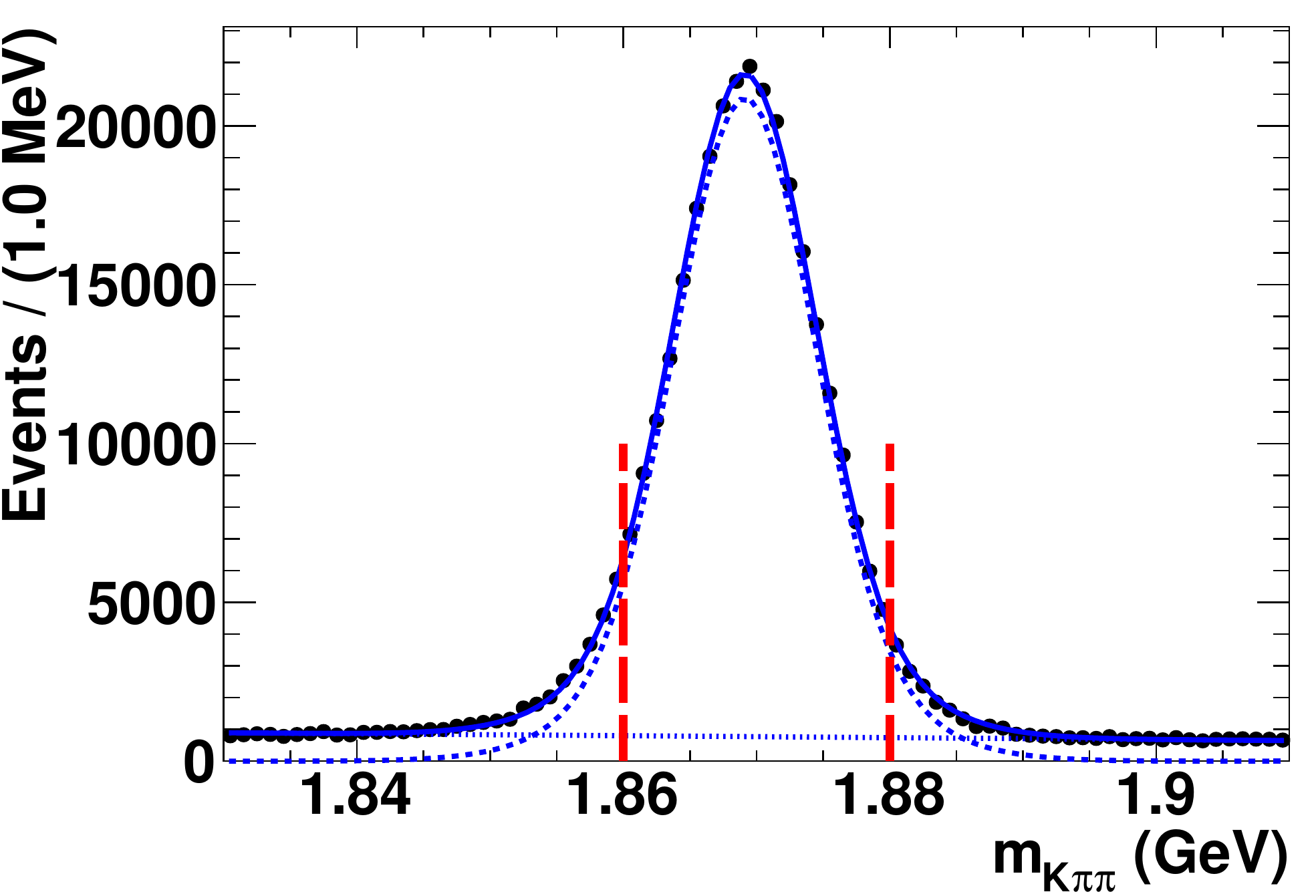}
\caption{(color online) The reconstructed $\Dp$ mass distribution 
of real data, after all $\Dstarp$ selection criteria except for the $\Dp$ mass requirement,  
which is marked by the two vertical dashed lines. 
The result of the fit described in the text is superimposed
(solid line), together with the background (dotted line) and signal 
(dashed line) components. 
}
\label{fig:DpMass}
\end{figure}

Decay candidates $\Dp \to \Km\pip\pip$ are formed from well-measured tracks with
kaon or pion particle identification and with a 
$\Km\pip\pip$ invariant mass $m_{K\pi\pi}$ within
1.86 and 1.88~\gev (approximately $\pm 2\sigma$ 
around the nominal \Dp mass~\cite{pdg}). 
This reduces background from random combinations of tracks, 
especially from $\Dstar\to D\pi_s^0$ decays with a correctly
reconstructed $\pi_s^0$, 
which will also peak in the 
 signal region of the $\Delta m$ distribution. 
As in Ref.~\cite{dstd0prl}, we reject candidates with any $\Dp$ daughter track for which the cosine of the polar angle 
measured in the laboratory frame 
 $\cos \theta_t$ is above 0.89; this criterion reduces the final sample by approximately 10\%. 
To further suppress peaking background events, 
we use a likelihood variable to select $\Dp$ candidates,
based on measured decay vertex separation from the primary vertex, and on 
Dalitz-plot position.
This likelihood criterion rejects about 70\% 
of background events with incorrectly reconstructed $\Dp$, while retaining
about 77\% of 
signal events. 
Figure~\ref{fig:DpMass} shows the $m_{K\pi\pi}$ distribution for 
data events passing all selection criteria except for the requirement on $m_{K\pi\pi}$.  
For illustrative purposes, we fit the $m_{K\pi\pi}$ distribution by modeling the $\Dp$ signal
with a sum of two Gaussian functions sharing a common mean, and random background events with a linear function. 
After all selection criteria, the fraction of candidates with a correctly reconstructed $\Dp$, 
as estimated from the $m_{K\pi\pi}$ fit, is about 95\%. 

\begin{figure*}[ht]
\centering
\subfigure{\label{fig:rdfits-a}
\includegraphics[width=0.47\linewidth]{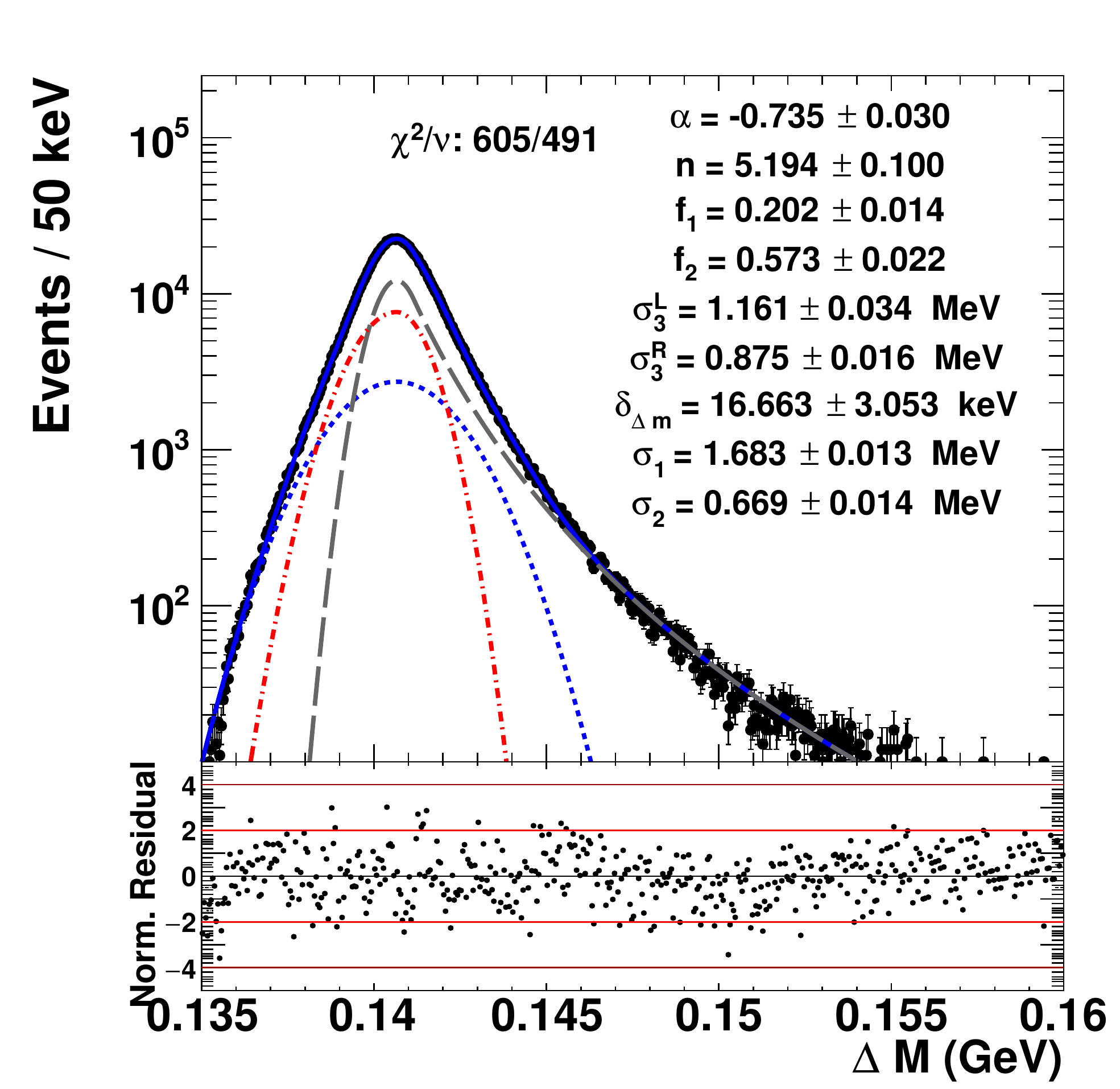}
\putat{-190}{+198}{\huge (a)}
\putat{-192}{+183}{\large MC}
}
\subfigure{\label{fig:rdfits-b}
\includegraphics[width=0.47\linewidth]{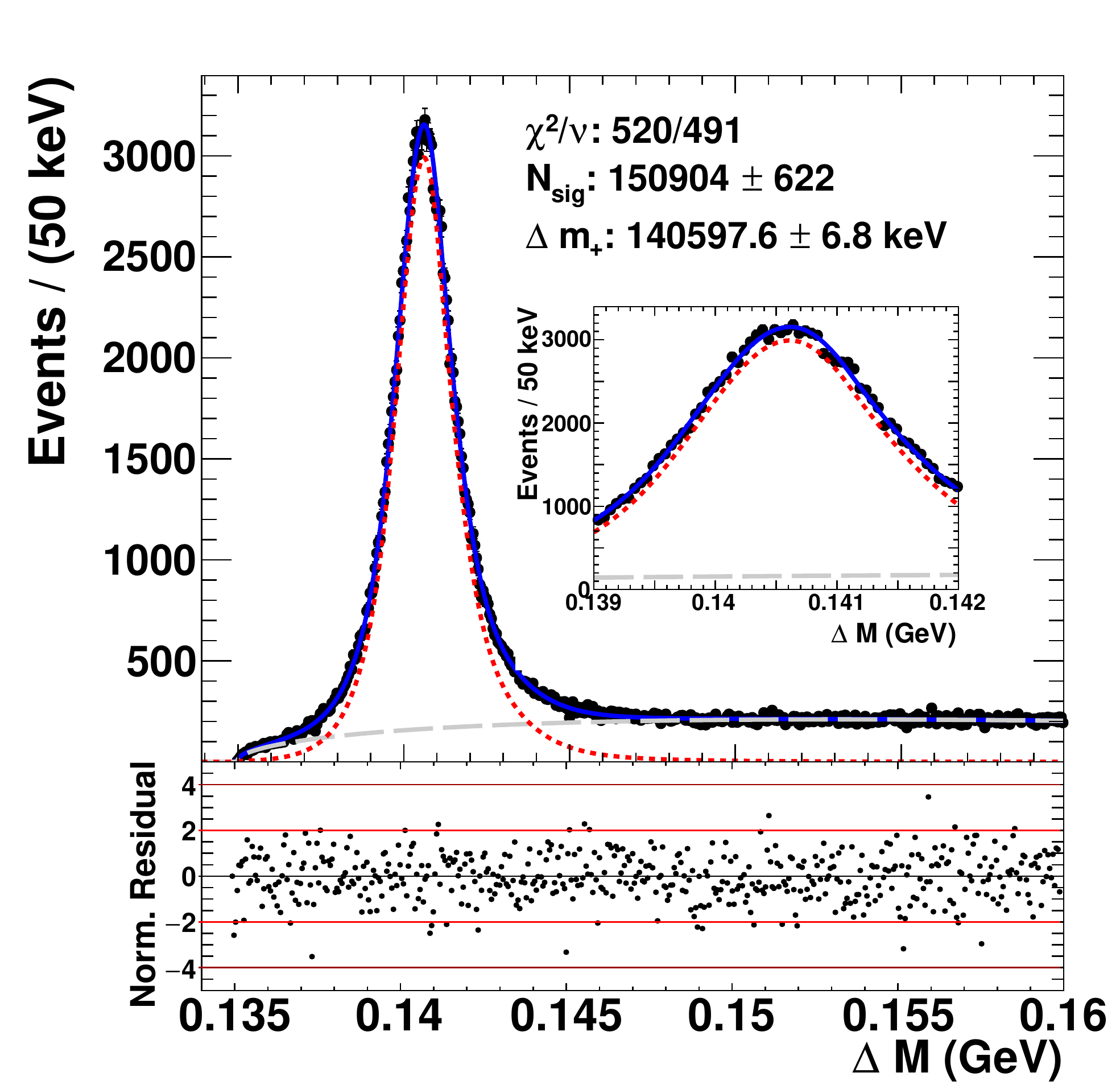} 
\putat{-190}{+198}{\huge (b)}
\putat{-192}{+183}{\large Data}
}
\caption{(color online) Left: $\Delta m$ fit to  correctly reconstructed signal MC events. 
    Shown are the total fit (blue solid line), Crystal Ball function (gray long-dashed line), Gaussian (blue short-dashed line), and  
    two-piece normal distribution function
    (red dash-dotted line). 
    The fitted signal shape parameters defined in Eq.~\ref{eq:sigpdf} are also shown in the text box. 
Right: $\Delta m$ fit to real data. Shown are the total fit (blue solid line), signal PDF (magenta short-dashed line), and background PDF (gray long-dashed line). 
The inset shows the fit around the peak region. 
The $\Delta m_+$ central value from the fit is later corrected 
by the estimated fit bias.
Normalized residuals shown underneath both fit plots are defined as $\left(N_{\text{observed}} - N_{\text{predicted}}\right)/\sqrt{N_{\text{predicted}}}$.}
\label{fig:rdfits}
\end{figure*}

The value of $\Delta m_+$ is obtained from a fit to the $\Delta m$ 
distribution in a two-step procedure as illustrated in Fig.~\ref{fig:rdfits} (a)
 and (b). First, we model the $\Delta m$ resolution function 
 by fitting the $\Delta m$ distribution for correctly reconstructed 
signal MC events 
using an 
empirically-motivated sum of three Gaussian or Gaussian-like probability density 
functions (PDFs): 
\begin{multline}
{\cal S}(\Delta m) = f_1 {\rm G}(\Delta m; \Delta m_++\delta_{\Delta m_+}, \sigma_1)\\ 
        + (1-f_1)\left[f_2 {\rm CB}(\Delta m;\Delta m_++\delta_{\Delta m_+}, \sigma_2, \alpha, n)\right. \\
        \left. + (1-f_2) {\rm BfG}(\Delta m;\Delta m_++\delta_{\Delta m_+}, \sigma^{L}_3,\sigma^{R}_3)\right]\,,\label{eq:sigpdf}
\end{multline}
where $f_1$ and $f_2$ give 
the fractions for the composite PDFs of G (Gaussian), CB 
(Crystal Ball~\footnote{M.~J.~Oreglia, Ph.D. thesis, Stanford University Report No. SLAC-R-236, 1980; J.~E.~Gaiser, Ph.D. thesis, Stanford University Report No. SLAC-R-255, 1982; T. Skwarnicki, Ph.D. thesis, Cracow Institute of Nuclear Physics Report No. DESY-F31-86-02, 1986.}, 
with $\alpha$ and $n$ as two parameters to model the high mass tail), and BfG (
a two-piece normal distribution 
with widths $\sigma^{L}_3$ and $\sigma^{R}_3$ on the left and right of $(\Delta m_++\delta_{\Delta m_+})$, respectively).  
The sum $(\Delta m_++\delta_{\Delta m_+})$ is therefore the common peak position of the three PDFs. 
In the fit to the high-statistics MC sample (Fig.~\ref{fig:rdfits}(a)), 
$\Delta m_+$ is fixed at the generated value of 140.636~\mev, and 
$\delta_{\Delta m_+}$ is a measure of the possible bias induced by 
our event selection procedure, or the chosen form for the resolution
function. 
The fitted functional distribution provides a reasonably good description of 
the data (with $  \chi^2 / \nu = 605 / 491 $ for a sample more than 7 times larger than the data). 
The fit gives $\delta_{\Delta m_+} = (+16.6\pm2.5)$~\kev, with the uncertainty 
from the limited size of our MC sample.
The fit results for the shape parameters are shown in Fig.~\ref{fig:rdfits}(a); 
and the full-width at half maximum (FWHM) of the resolution function is found to be about 2.1~\mev, 
which is mainly due to  
the resolution of the $\pi^0_s$.

The second step (Fig.~\ref{fig:rdfits}(b)) is an unbinned maximum-likelihood fit to real data 
using the PDF from the first step to model signal, and a threshold function to model the combinatorial background~\cite{Albrecht:1990cs}:

\begin{equation}
T(\Delta m; \kappa) = \Delta m \sqrt{u} \exp(\kappa\cdot u)\,, 
\end{equation}
where $u = (\Delta m/ m_{\rm endpt})^2-1$, and $\kappa$ is the slope parameter 
which is allowed to vary in the fit. 
We fix the end point $m_{\rm endpt}$ at the nominal $\piz$ mass~\cite{pdg} as the physical limit of $\Delta m$. 
In the data fit, we fix the bias $\delta_{\Delta m_+}$, fractions $f_{1,2}$, and CB tail parameters to the MC values from the first step, 
while allowing the widths $\sigma_{1,2,3}$ to be free in the fit to allow for differences between MC simulation and data. 
Figure~\ref{fig:rdfits}(b) presents the data and the fit, with the normalized residuals showing good data and fit agreement. 
There are $150\,904 \pm622$ signal events, 
the observed FWHM of the signal shape is about 2.0~\mev, 
and
we determine  
$\Delta m_+ = (140\,597.6 \pm 6.8)$~\kev, 
where the uncertainty is statistical only ($\sigma_{\rm stat}$). 
A bias correction to this result will be discussed later.

We estimate systematic uncertainties on $\Delta m_+$ from a variety of sources.
Separately, we study the $\Delta m_+$ dependence on the $D^{*+}$ laboratory momentum $p_{\rm lab}$, on the cosine of $D^{*+}$ laboratory polar angle $\cos\theta$, on the $D^{*+}$ laboratory azimuthal angle $\phi$, on $m_{K\pi\pi}$, and on the diphoton opening angle $\theta_{\gamma\gamma}$ from $\piz\to \gamma\gamma$, by collecting fit results for $\Delta m_+$ in 10 subsets of data with 
roughly equal statistics for each parameter.
Furthermore, 
we divide our data into four disjoint subsets of data-taking periods.
For the data fit in each subset, the value of $\delta_{\Delta m_+}$
is determined separately from signal MC events with the same event selection criteria as for that subset. 
This is  
 meant to expose possible detector response effects
that have not been modeled in the simulation.
We search for variations larger than those expected from statistical fluctuations based on a method similar to the PDG scale factor~\cite{prdversion, pdg}. 
If the fit results from a given dependence study are compatible with a constant value, in the sense that $\chi^2/\nu<1$, 
where $\nu$ is the number of degrees of freedom,
we assign no systematic uncertainty. In the case that  $\chi^2/\nu>1$, we ascribe an uncertainty of $\sigma_{\rm sys} = \sigma_{\rm stat}\sqrt{\chi^2/\nu-1}$ to account for unidentified detector effects. 
We observe $\chi^2/\nu > 1$ in the cases of $p_{\rm lab}$,  
$\cos\theta$ and $\theta_{\gamma\gamma}$ (shown  
in~\footnote{Additional plots are available through EPAPS Document No. E-PRLTAO-XX-XXXXX.
    For more information on EPAPS, see http://www.aip.org/pubservs/epaps.html.}). 
Systematic uncertainties of 
5.0~\kev, 6.9~\kev, and 6.1~\kev 
are assigned for the 
$ D^{*+} \ p_{\rm lab} $, $ D^{*+} \  \cos \theta $, and 
$ \theta_{\gamma\gamma} $ dependences, respectively, for which the $ p $-values for the null hypotheses are 0.12, 0.03, and 0.06.
The $ p $-values for the variations with $ D^{*+} $ azimuthal angle and $ D^+ $ mass are 0.99 and 0.47, 
and no systematic uncertainties are assigned for these observations.

The five signal shape parameters $\alpha$, $n$, 
$f_{1,2}$, and $\delta_{\Delta_{m_+}}$, determined 
from the fit to signal MC events (Fig.~\ref{fig:rdfits} (a)), 
possess statistical uncertainties that are highly correlated.
We account for their uncertainties and correlations by
producing 100 sets of correlated random numbers of signal shape 
parameters based on
the central values and the covariance matrix from the fit to signal 
MC events. Then for each set, we rerun the data fit by fixing $\alpha$, $n$, 
$f_{1,2}$, and $\delta_{\Delta_{m_+}}$ to the corresponding 
random numbers in the set. 
The distribution of the 100 fit values for
$\Delta_{m_+}$ has a root mean square of 2.1~\kev which is taken as
systematic uncertainty for the signal shape parameters.

To test whether our fit  
procedure introduces a
bias on $\Delta m_+$, 
we generate an ensemble of data sets with signal and background events 
generated from appropriately normalized PDFs based on our nominal data fit. 
The data sets are then fitted with exactly the same fit model as for real data (``pure pseudoexperiment''). 
By performing 500 pseudoexperiments, 
we collect 
$\Delta m_+$ pulls, defined as the differences of fitted and input values normalized by the fitted errors. The 
mean of the pulls is 
$-(50\pm4)\%$, 
while the root mean square is consistent with being unity.
We thus correct for the bias in our fit model
by adding 
{$50\% \times \sigma_{\rm stat} = 3.4$~\kev}
to the fit value of $\Delta m_+$ from the data, and assign a systematic uncertainty equal to half this bias correction 
({$1.7$~\kev}).
We perform another type of pseudoexperiment by fitting
to ensembles of data sets where signal and background events are
produced by randomly sampling the corresponding MC 
events. 
Background events from decays such as  
$ D^{*+} \to D^+ \pi^0 $ with 
$ D^+ \to \pi^- \pi^+ \pi^+ \pi^0 $ misreconstructed as $ K^- \pi^+ \pi^+ $ 
produce small peaks in the signal region, but 
the fit does not account for them explicitly. 
The collected pulls show a mean fit bias consistent with that 
found in our pure pseudoexperiments, and we assign no additional systematic 
uncertainty related to peaking backgrounds.

To account for the systematic uncertainty due to
    imperfect photon energy 
    simulation and calibration in the MC, 
    we rescale photon energies in 
    signal MC events
    by $+0.3\%$ and $-0.3\%$, and take the larger of the two variations in the $\Delta m$ peak position,
    7.0\kev, as the corresponding systematic uncertainty.    
The values $ \pm 0.3\% $  correspond to the difference between MC and
data $ \piz $ mass peak positions after the nominal MC neutral energy corrections
are applied.
Because the MC and data $m_{\gamma\gamma}$ distribution shapes differ,
aligning the peak positions does not produce equal mean values.
We also account for the associated
uncertainties on the $\piz$ momentum rescaling factors due to the limited size of our MC sample, 
and find the related systematic uncertainty to be 0.5~\kev.

Besides the systematic studies, we also perform a series of consistency checks that are not used to assess systematics 
but rather to reassure us that the experimental approach and fitting technique behave reasonably. 
We vary the upper limit of the $\Delta m$ fit range from its default position of 0.160~\gev to a series
of values between 0.158 and 0.168~\gev. Also, we vary the selection criteria 
on the invariant masses $m_{K\pi\pi}$ and $m_{\gamma\gamma}$, as well as the Dalitz-plot based likelihood. 
The resulting fit values of $\Delta m_+$ from all these checks are consistent. 

\begin{table}
\centering
\caption{\label{tab:summary} Assigned systematic errors from all considered sources.}
\begin{tabular}{|l|c|}\hline
Source                                       & $\Delta m_+$ systematic [keV]  \\ \hline
Fit bias             & 1.7 \\
   $\Dstarp$ $p_{\rm lab}$ dependence    &    5.0   \\ 
    $\Dstarp$ $\cos\theta$ dependence    &    6.9      \\ 
$\Dstarp$ $\phi$ dependence    &    0.0      \\ 
$m(D^+_{\rm reco})$ dependence        &    0.0         \\ 
    Diphoton opening angle dependence        &   6.1         \\ 
Run period dependence               &    0.0       \\ 
Signal model parametrization        &    2.1     \\
EMC calibration      &    7.0            \\ 
MC $\piz$ momentum rescaling      &    0.5            \\ 
    \hline
Total  & \finalsys \\ \hline
\end{tabular}
\label{table:syswithcorr}
\end{table}

    All systematic uncertainties of $\Delta_{m_+}$ are
summarized in Table ~\ref{table:syswithcorr}; adding them in quadrature leads to a total
of 12.9~\kev. After adding the fit bias of 3.4 keV,
our final result is 
    $\Delta m_+ \equiv m(\Dstarp) - m(\Dp) = (140\,601.0 \pm 6.8[{\rm stat}]  
    \pm \finalsys[{\rm syst}])$~\kev. 
    This result is consistent with the current world average of ($140.66 \pm 0.08 $)~\mev, 
and about five times more precise. 
    Combining with the \babar measurement of $\Delta m_0 = (145\,425.9 \pm 0.5[{\rm stat}] \pm 1.8[{\rm syst}])$~\kev based
on the same data set, we obtain the $D$ meson mass difference of 
$\Delta m_D = (4\,824.9 \pm 6.8[{\rm stat}]  
\pm \finalsys[{\rm syst}])$~\kev. 
This result is, as for $ \Delta m_+ $, about a factor of five 
more precise than the current world average, ($4.77\pm0.08$)~\mev. 
Adding the statistical and systematic uncertainties in quadrature, 
$ \Delta m_D = (4 \, 824.9 \pm 14.6)$~\kev.
This can be compared with the corresponding values for the pion and kaon systems, 
$ \Delta m_{\pi} = (4\, 539.6 \pm 0.5) $~\kev and $ \Delta m_K = (- 3\, 934 \pm 20 ) $~\kev~\cite{pdg}.

We are grateful for the excellent luminosity and machine conditions
provided by our \pep2\ colleagues, 
and for the substantial dedicated effort from
the computing organizations that support \babar.
The collaborating institutions wish to thank 
SLAC for its support and kind hospitality. 
This work is supported by
DOE
and NSF (USA),
NSERC (Canada),
CEA and
CNRS-IN2P3
(France),
BMBF and DFG
(Germany),
INFN (Italy),
FOM (The Netherlands),
NFR (Norway),
MES (Russia),
MINECO (Spain),
STFC (United Kingdom),
BSF (USA-Israel). 
Individuals have received support from the
Marie Curie EIF (European Union)
and the A.~P.~Sloan Foundation (USA).


\bibliography{myrefs}
\onecolumngrid
\newpage

\section{EPAPS Material}

\begin{figure}[h!]
\centering
\subfigure{
\includegraphics[width=0.48\textwidth]{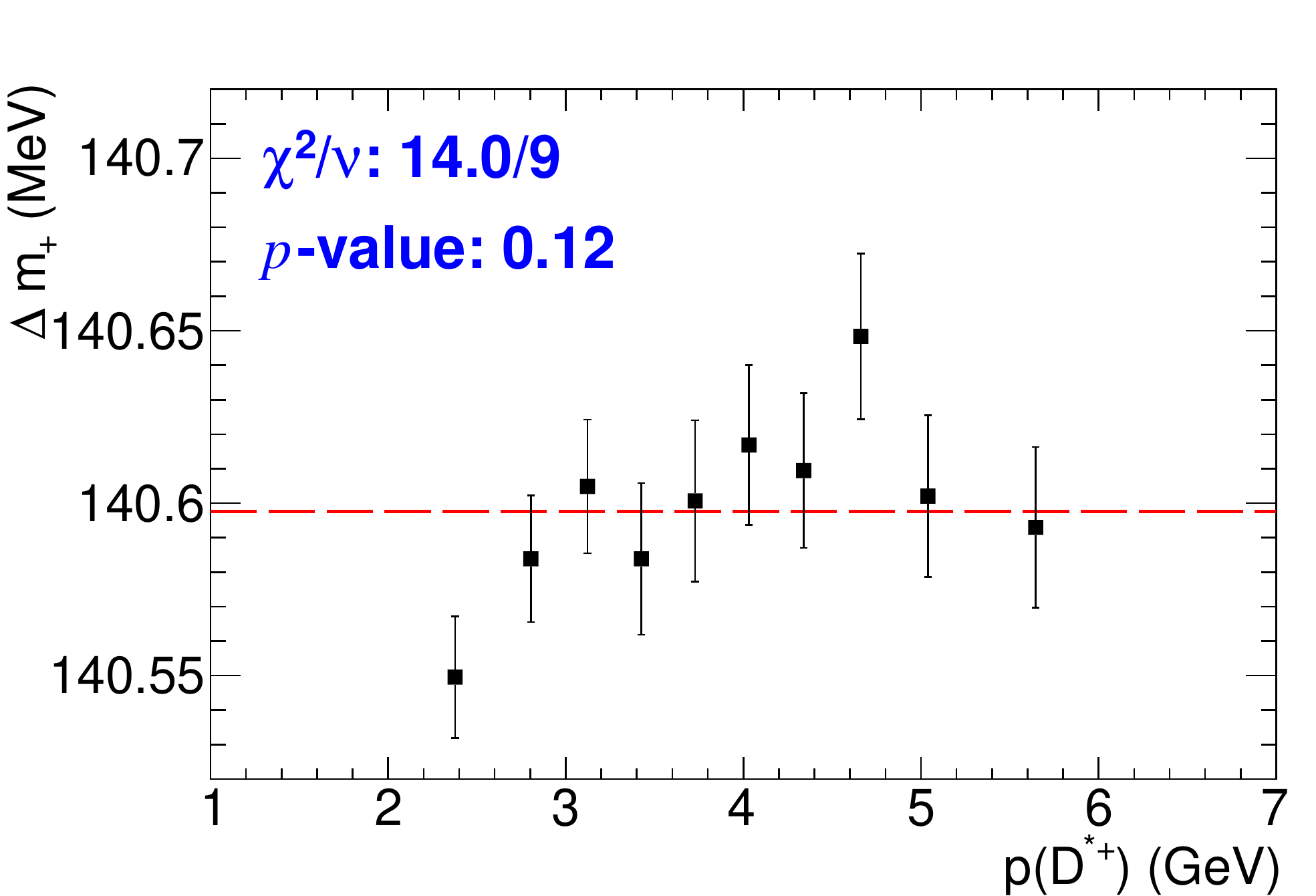} 
\putat{-50}{+130}{\huge (a)}}
\subfigure{
\includegraphics[width=0.48\textwidth]{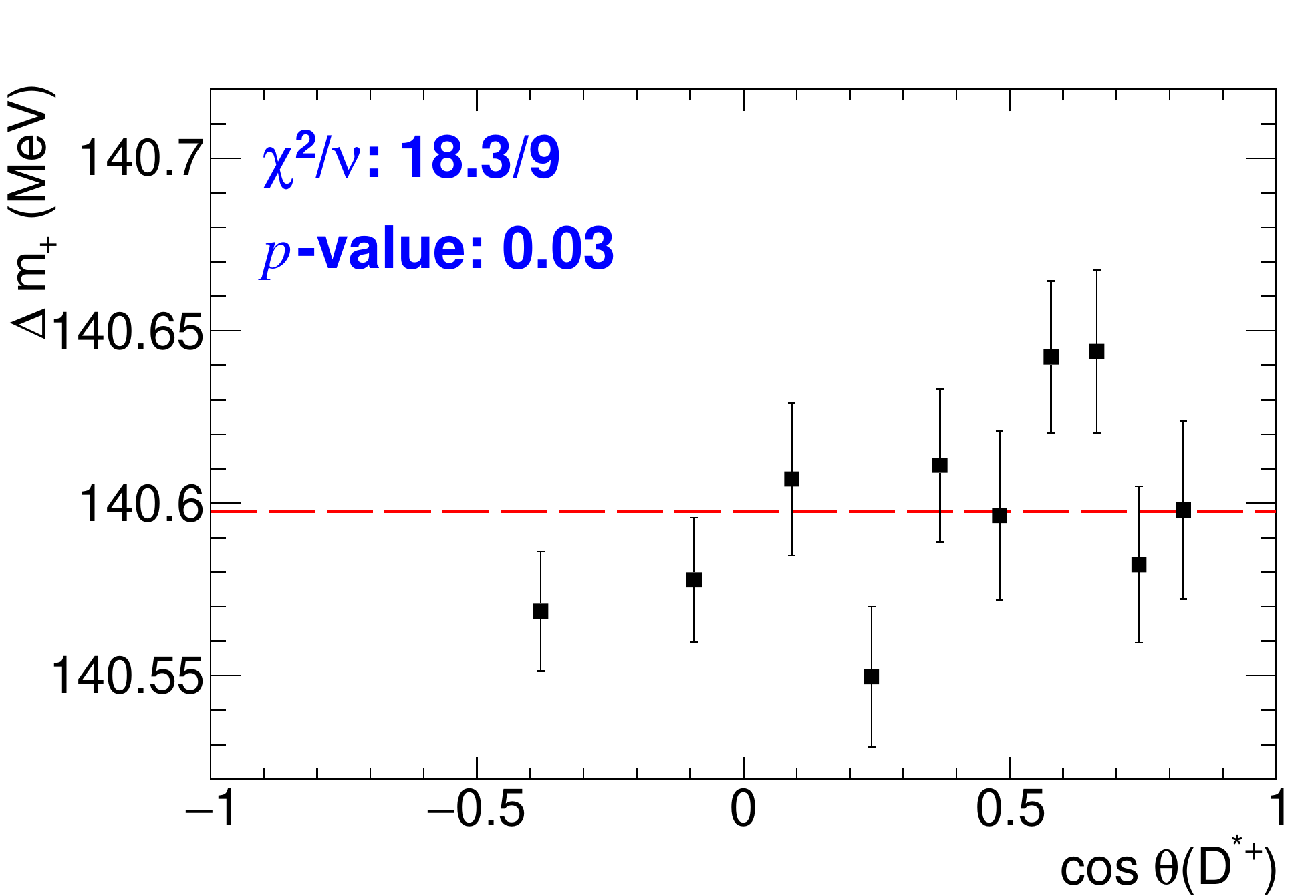} 
\putat{-50}{+130}{\huge (b)}}
\subfigure{
\includegraphics[width=0.48\textwidth]{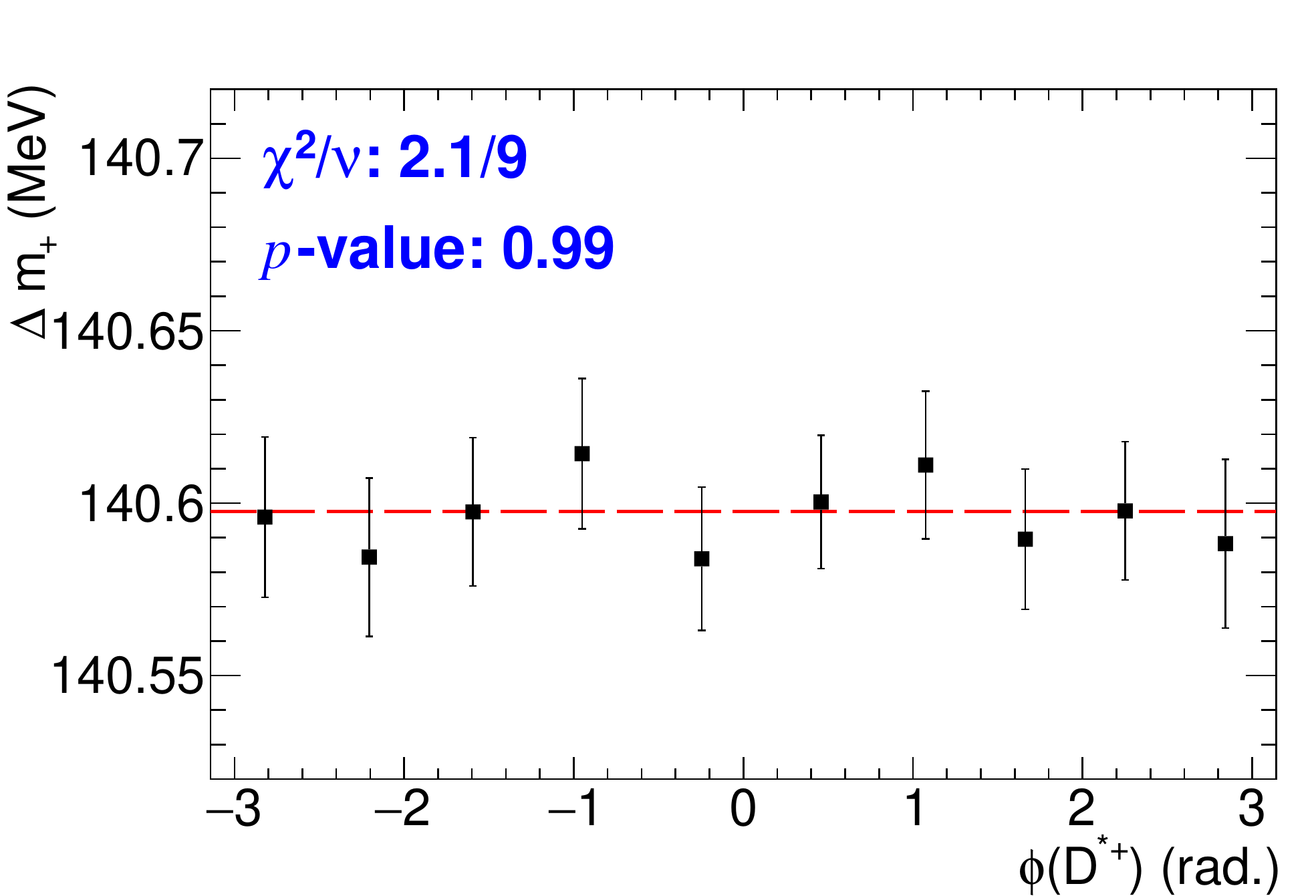} 
\putat{-50}{+130}{\huge (c)}}
\subfigure{
\includegraphics[width=0.48\textwidth]{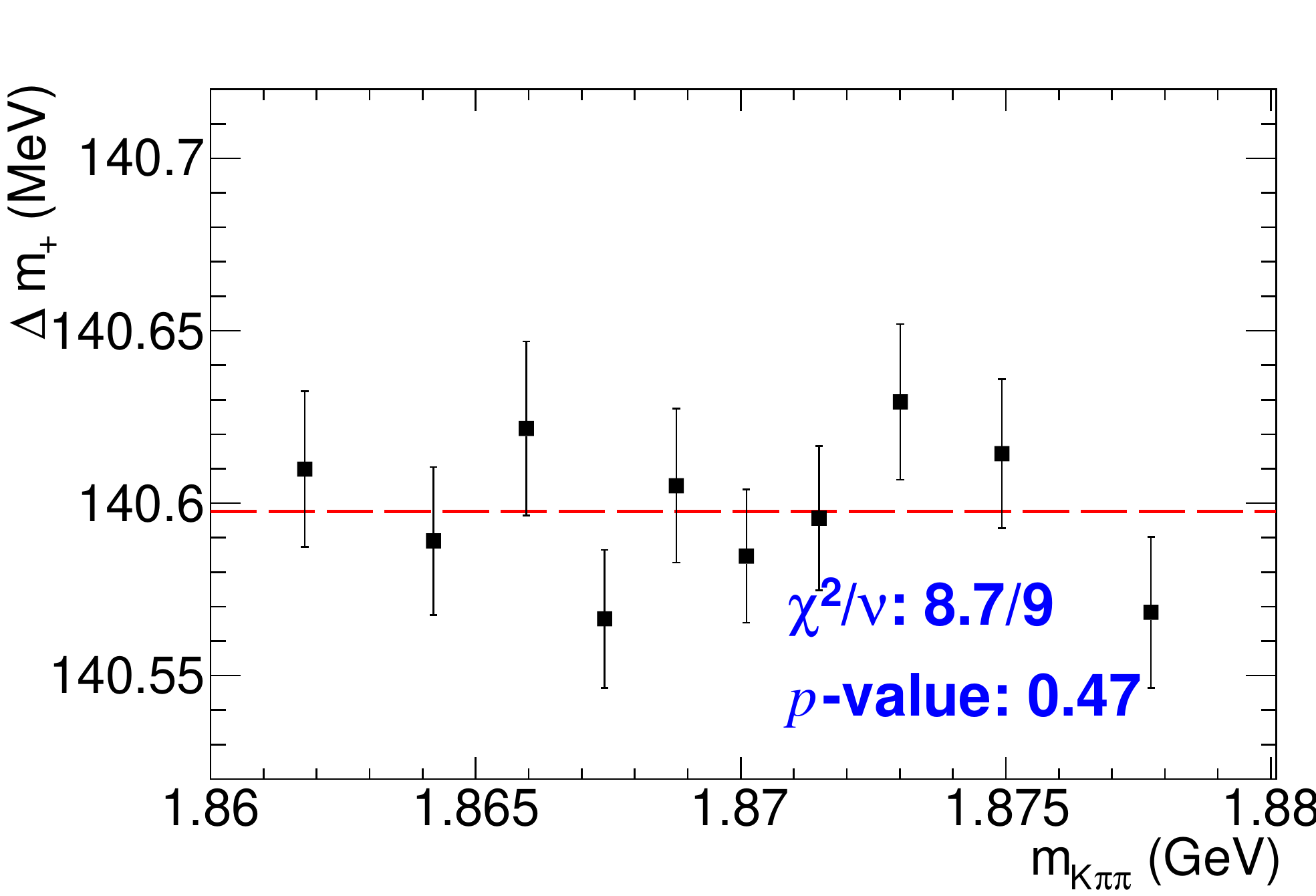} 
\putat{-50}{+130}{\huge (d)}}
\subfigure{
\includegraphics[width=0.48\textwidth]{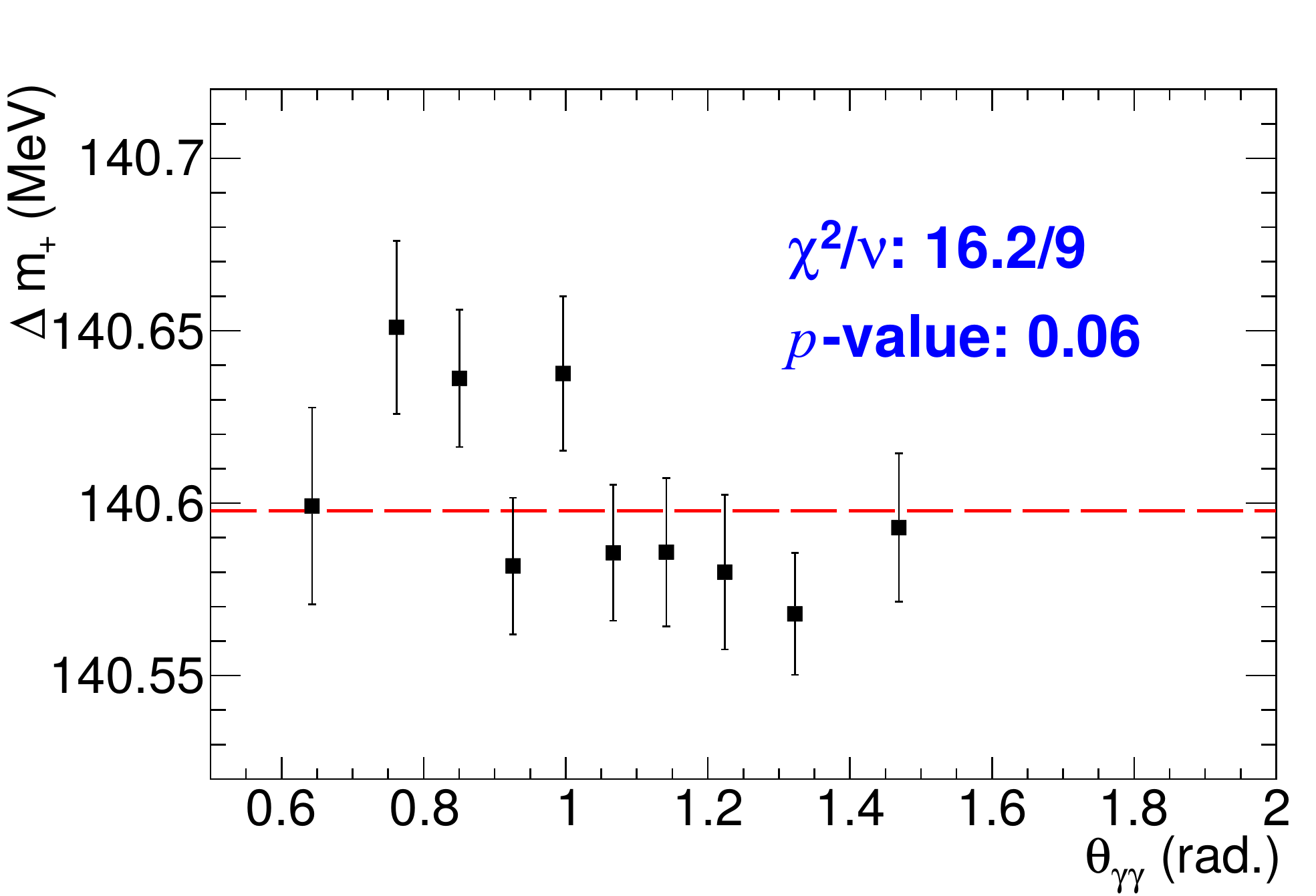} 
\putat{-50}{+130}{\huge (e)}}
\caption{\label{fig:PDGmethods}(color online) $\Delta m_+$ measurements 
    as functions of different $\Dstarp$ properties:
    (a) momentum magnitude $p(\Dstarp)$, 
(b) polar angle $\cos\theta$, 
    (c) azimuthal angle $\phi$, (d) $m_{K\pi\pi}$, 
    and (e) $\piz$ opening angle $\theta_{\gamma\gamma}$.
    In each plot we fit the results with a constant and the fitted $\chi^2/\nu$ with associated $p$-value is shown in each plot.
The red dashed lines mark the $\Delta m_+$ central value from our nominal fit.
}
\end{figure}

\end{document}